\documentclass[
    ,final            % use final for the camera ready runs
  ]
  {aipproc}

\layoutstyle{8x11double}
\usepackage{amssymb}

\newcommand{\Teff}{T_{\rm eff}}
\newcommand{\ThetaB}{\Theta_B}
\newcommand{\vecB}{\mathbf B}
\newcommand{\zg}{z_g}
\newcommand{\Rinfty}{R^{\infty}}
\newcommand{\xmm}{\textit{XMM-Newton}}
\newcommand{\rxj}{RX~J1856.5$-$3754}
\newcommand{\onee}{1E~1207.4$-$5209}

\begin{document}

%%%%%%%%%%%%%%%%%%%%%%%%%%%%%%%%%%%%%%%%%%%%%%%%%%%%%%%%%
\title{Modeling Phase-resolved Observations of the Surfaces of Magnetic
Neutron Stars}


\author{Wynn C.~G. Ho}{
  address={Harvard-Smithsonian Center for Astrophysics, 60 Garden St.,
Cambridge, MA, 02138, USA}
}
\author{Kaya Mori}{
  address={Department of Astronomy and Astrophysics, University of Toronto,
50 St. George Street, Toronto, Ontario, M5S 3H4, Canada}
}

%%%%%%%%%%%%%%%%%%%%%%%%%%%%%%%%%%%%%%%%%%%%%%%%%%%%%%%%%
\begin{abstract}
Recent observations by \xmm\ detected rotational pulsations in the total
brightness and spectrum of several neutron stars.  To properly
interpret the data, accurate modeling of neutron star emission is necessary.
Detailed analysis of the shape and strength of the rotational variations
allows a measurement of the surface composition and magnetic field,
as well as constrains the nuclear equation of state.
We discuss our models of the spectra and light curves of two of the most
observed neutron stars, \rxj\ and \onee, and discuss some implications of
our results and the direction of future work.
\end{abstract}

\maketitle

%%%%%%%%%%%%%%%%%%%%%%%%%%%%%%%%%%%%%%%%%%%%%%%%%%%%%%%%%
\section{Introduction \label{sec:intro}}

Thermal radiation from the surface of neutron stars (NSs) can provide
invaluable information on the physical properties and evolution of NSs.
NS properties, such as the mass $M$ and radius $R$,
in turn depend on the poorly constrained physics
of the stellar interior, such as the nuclear equation of state (EOS) and
quark and superfluid/superconducting properties at supra-nuclear densities.
Many NSs are also known to possess strong magnetic fields
($B \sim 10^{12}-10^{13}$~G), with some well above the quantum critical
value ($B\gg B_{\rm Q}\equiv 4.4\times 10^{13}$~G).
 
The observed thermal radiation originates in a thin atmospheric
layer (with scale height $\sim 1$~cm) that covers the stellar surface.
To properly interpret the observations of NS surface emission and to
provide accurate constraints on their physical properties, it is
important to understand in detail the radiative behavior of NS
atmospheres in the presence of strong magnetic fields
(see \citep{pavlovetal95,holai01,zavlinpavlov02,holai03,zavlin07},
for more detailed references on observations and on previous works
in NS atmosphere modeling).
The properties of the atmosphere, such as the chemical composition,
EOS, and radiative opacities, directly determine the characteristics
of the observed spectrum.
While the surface composition of the NS is unknown,
a great simplification arises due to the efficient gravitational
separation of light and heavy elements \citep{alcockillarionov80}.
A pure hydrogen atmosphere is expected even if a small
amount of accretion/fallback occurs after NS formation;
the total mass of hydrogen needed to form an optically thick atmosphere
can be less than $\sim 10^{16}$~g.
On the other hand, a heavy element atmosphere may be possible if no
accretion takes place.

The strong magnetic fields present in NS atmospheres significantly
increase the binding energies of atoms, molecules, and other bound states
(see \citep{lai01}, for a review).  Abundances of these bound states can
be appreciable in the atmospheres of cold NSs (i.e., those with surface
temperature $T\lesssim 10^6$~K; \citep{laisalpeter97,potekhinetal99}).
In addition, the presence of a magnetic field causes emission to be
anisotropic and polarized; this must be taken into account
when developing radiative transfer codes.
The most comprehensive early studies of magnetic NS
atmospheres focused on a fully ionized hydrogen plasma and
moderate field strengths ($B\sim 10^{12}-10^{13}$~G;
\citep{miller92,shibanovetal92,pavlovetal94,zaneetal00}).
These models are expected to be valid only for relatively high temperatures.
More recently, atmosphere models in the ultra-strong field
($B\gtrsim 10^{14}$~G) and relevant temperature regimes have
been presented (\citep{ozel01,zaneetal01,holai03,lloyd03,vanadelsberglai06};
see also \citep{bezchastnovetal96,bulikmiller97}, for early work), and all
of these rely on the assumption of a fully ionized hydrogen composition.
Magnetized non-hydrogen atmospheres have been studied by
\citep{miller92,rajagopaletal97}, but because of the complexity of the
atomic physics, the models were necessarily crude
(see \citep{moriho07}, for more details).
Only recently has self-consistent atmosphere models
\citep{hoetal03,potekhinetal04,moriho07} using the latest EOS and opacities
for partially ionized hydrogen \citep{potekhinchabrier03,potekhinchabrier04}
and mid-$Z$ elements \citep{morihailey02,morihailey06} been constructed.

The atmosphere models discussed above only describe emission from a
local patch of the stellar surface.
By taking into account surface magnetic field $\vecB$ and temperature $T$
distributions, we can construct more physically correct models of emission
from NSs.
However, these spectra from the whole NS surface are necessarily
model-dependent, as the $\vecB$ and $T$ distributions are unknown.
Nevertheless, detailed comparisons of the models with rotation phase-resolved
observations is a powerful tool to study NSs, e.g., spectral features
that vary with phase are essential to disentangling magnetic
field effects from other parameters and to probe the magnetic field
geometry on the surface of the star.
Indeed there have been recent works attempting to fit magnetic
atmosphere spectra to observations of NSs (see \citep{ho07}, and
references therein).
Here we describe some of the details and observational applications
of our work.

%%%%%%%%%%%%%%%%%%%%%%%%%%%%%%%%%%%%%%%%%%%%%%%%%%%%%%%%%
\section{Model for Neutron Star Surface Emission \label{sec:model}}

In order to construct models of emission from the entire NS surface,
we first build atmosphere models for a given effective temperature
$\Teff$ and magnetic field strength $B$ and direction $\ThetaB$
relative to the surface normal;
these (local) models describe a patch of the NS surface.
In the presence of magnetic fields $B\gtrsim 10^{12}$~G, radiation propagates
in two photon polarization modes (see, e.g., \citep{meszaros92});
therefore, the atmosphere models are obtained by solving the radiative
transfer equations for the two coupled polarization modes
(see \citep{holai01,hoetal03,hoetal07,potekhinetal04}, for details on the
construction of the atmosphere models).
In addition to $\vecB$ and $\Teff$,
the atmosphere models have a dependence, through hydrostatic
balance, on the surface gravity $g$~$[=(1+\zg)GM/R^2]$,
where the gravitational redshift $\zg$ is given by
$(1+\zg)=(1-2GM/Rc^2)^{-1/2}$;
however, the resulting spectra do not vary significantly using different
values of $g$ around $2\times 10^{14}$~cm~s$^{-2}$ \citep{pavlovetal95}.

Next, the entire NS surface is divided into regions with different
$\vecB$ and $\Teff$.  Relatively simple surface distributions of
$\vecB$ and $\Teff$ are adopted: we assume the surface is symmetric
(in $\vecB$ and $\Teff$) about the magnetic equator and divide the
hemisphere into several magnetic colatitudinal regions.
An example parametrization is given in Table~\ref{tab:nssurf};
note that the magnetic field distribution is roughly dipolar.
Emission from any point within a colatitudinal region is given by the
atmosphere model for that region.

%-------------------------------------------
\begin{table}[htb]
\begin{tabular}{c c c c}
\hline
\tablehead{1}{c}{c}{magnetic colatitude}
 & \tablehead{1}{c}{c}{$B$}
 & \tablehead{1}{c}{c}{$\ThetaB$}
 & \tablehead{1}{c}{c}{$\Teff$} \\
 (deg) & ($10^{12}$~G) & (deg) & ($10^6$~K) \\
\hline
0$-$20 & 6 & 0 & 7 \\
20$-$50 & 5 & 30 & 6 \\
50$-$70 & 4 & 60 & 5 \\
70$-$90 & 3 & 90 & 4 \\
\hline
\end{tabular}
\caption{Neutron Star Surface Parametrization \label{tab:nssurf}}
\end{table}
%-------------------------------------------

Finally, the spectra from the entire NS surface is computed by
summing over the emission from the different regions for a given
rotation phase (see \citep{pavlovzavlin00,ho07}, for details).
The observed emission depends on two angles ($\alpha,\zeta$):
$\alpha$ is the angle between the rotation and magnetic
axes and $\zeta$ is the angle between the rotation axis and the
direction to the observer.
We also account for the bending of the path of light due to gravity,
which causes more of the NS surface to be visible
($135^\circ$ and $115^\circ$ as compared to $90^\circ$ without
light-bending for $M=1.4M_\odot$ and $R=10$~km and 14~km, respectively;
see \citep{pechenicketal83,beloborodov02}).

%-------------------------------------------
\begin{figure}[htb]
\includegraphics[width=0.35\textwidth]{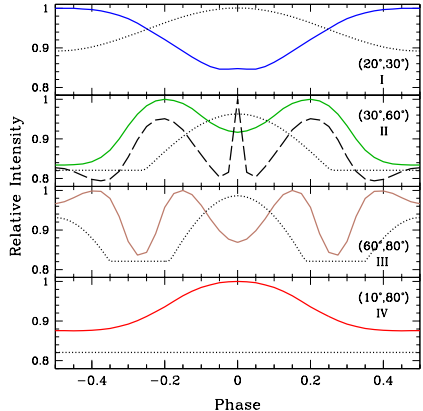}
\caption{
Light curves for different geometries ($\alpha,\zeta$):
class~I with (20$^\circ$,30$^\circ$), class~II with (30$^\circ$,60$^\circ$),
class~III with (60$^\circ$,80$^\circ$), and class~IV with
(10$^\circ$,80$^\circ$).
$\alpha$ is the angle between the spin and magnetic axes, and
$\zeta$ is the angle between the spin axis and the line-of-sight.
The four classes are defined in \citep{beloborodov02}.
The solid lines are for the magnetic model described in the text
[dashed line is for (50$^\circ$,50$^\circ$)],
while the dotted lines are analytic light curves (scaled arbitrarily
in amplitude) for isotropic emission from two antipodal hot spots
(see \citep{beloborodov02}).
\label{fig:nsmodelpulse}}
\end{figure}
%-------------------------------------------

Figure~\ref{fig:nsmodelpulse} shows the light curves, for various
geometries ($\alpha,\zeta$), of the NS model
using the parametrization given in Table~\ref{tab:nssurf}, $\zg=0.2$,
and $g=1.1\times 10^{14}$~cm~s$^{-2}$.
We also plot the analytic light curves from \citep{beloborodov02} for
isotropic emission from two antipodal hot spots
(see \citep{zavlinpavlov98,bogdanovetal07}, for
examples of pulse profiles from non-magnetic hydrogen atmosphere hot spots).
The classification scheme (for isotropically-emitting hot caps)
is defined in \citep{beloborodov02}:
(I) only the primary cap is visible, and the pulse profile is purely
sinusoidal with a single peak,
(II) the opposite cap is seen around pulse minimum due to relativistic
light-bending, which reduces the strength of the modulation,
(III) the primary cap is not seen during a segment of the rotation,
and (IV) both spots are seen at all phases and thus there is no modulation.

Several important features are evident from a comparison of magnetic
atmosphere emission to that of isotropic emission.
The angular-dependence of the radiation (or beam pattern)
manifests as a narrow ``pencil-beam'' along the direction of the magnetic
field and a broad ``fan-beam'' at intermediate angles
(see \citep{pavlovetal94,lloyd03},
for beam patterns and spectra at various $\ThetaB$).
As discussed in \citep{pavlovetal94}, the pencil-beam is the result of the
lower opacity at angles $\lesssim (E/E_B)^{1/2}$, where
$E_B=\hbar eB/m_{\rm e}c=11.6\,(B/10^{12}\mbox{ G})$~keV is the electron
cyclotron energy; the width of the pencil-beam is thus
$\sim (E/E_B)^{1/2}$, and the radiation is more strongly beamed at higher
magnetic fields.
This narrow beam is seen in the (50$^\circ$,50$^\circ$)-light curve plotted
in Figure~\ref{fig:nsmodelpulse}, which is the only instance shown that has
the observer's line-of-sight exactly crossing the magnetic cap and coinciding
with the peak of the isotropic emission.
Also evident is the fan-beam (most obvious in the light curves of classes
II and III), which occur on either side of the magnetic cap and can increase
the number of light-curve peaks.
Finally, the anisotropic beam pattern (combined with the surface
temperature variation) can produce an apparent phase shift compared to
isotropic emission and modulation when an isotropic beam pattern
suggests none (c.f. class IV).

%%%%%%%%%%%%%%%%%%%%%%%%%%%%%%%%%%%%%%%%%%%%%%%%%%%%%%%%%
\section{\rxj \label{sec:rxj1856}}

\begin{figure}[htb]
\includegraphics[width=0.4\textwidth]{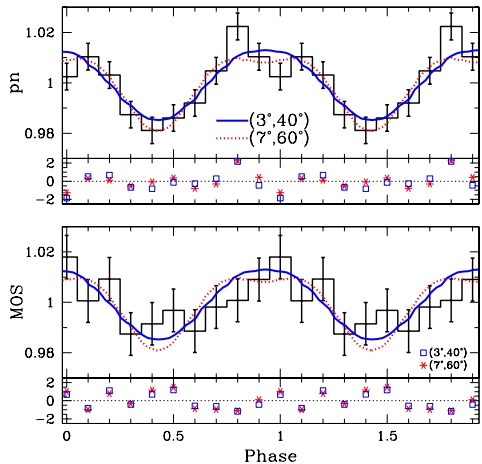}
\caption{ Energy-integrated (0.15$-$1.2~keV) light curves of \rxj.
Histograms are the \xmm\ EPIC-pn (top) and EPIC-MOS (bottom) observations
(see Fig.~1 of \citep{tiengomereghetti07}).
Solid and dotted lines are the models with ($3^\circ$,$40^\circ$) and
($7^\circ$,$60^\circ$), respectively, along with the fit deviations
[i.e., (data-model)/$\sigma$] in the corresponding lower panels.
Two rotation periods are shown for clarity.
\label{fig:pulse_rxj}}
\end{figure}

\rxj\ is one of the brightest, nearby isolated NSs \citep{kaplanvankerkwijk07},
and considerable observational resources have been devoted to its study.
Recently, \xmm\ observations
uncovered pulsations from \rxj\ with a period of 7~s
and an upper limit on the period derivative
\citep{tiengomereghetti07,mereghettitiengo07}.
The EPIC-pn and MOS light curves are shown in Figure~\ref{fig:pulse_rxj}.
Several characteristics of the observed light curves
suggest possible values of $\alpha$ and $\zeta$:
(1) the 1.6\%$\pm$0.2\% amplitude of the pulsations,
(2) a single peak (or two peaks close in phase, as may be the case
for the pn light curve) per rotation,
and (3) no significant energy-dependence for the single observation
and small energy-dependence for all
\xmm\ data \citep{ho07,tiengomereghetti07}.

%-------------------------------------------
%-------------------------------------------

\begin{figure}[htb]
\includegraphics[width=0.4\textwidth]{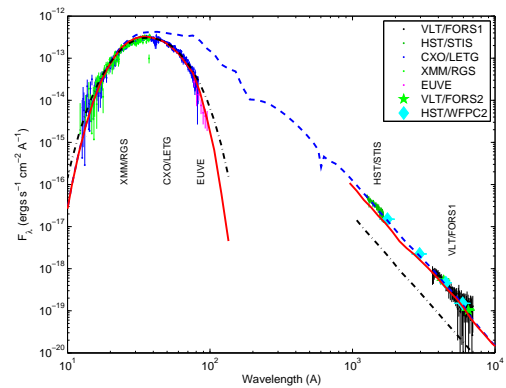}
\caption{ Spectrum of \rxj\ from optical to X-ray wavelengths.
Data points are observations taken from various sources;
error bars are 1$\sigma$ uncertainties.
The solid line is the absorbed (and redshifted by $\zg=0.22$) atmosphere
model spectrum with $B=4\times 10^{12}$~G, $\Teff = 5.3\times 10^5$~K,
and $\Rinfty=17$~km;
the dashed line is the unabsorbed atmosphere model spectrum.
The dash-dotted line is the (absorbed) blackbody fit to the X-ray spectrum
with $\Rinfty=5$~km.
Note that our atmosphere model underpredicts the optical flux by 15\%;
however, observational and model uncertainties here is $\sim 20\%$.
\label{fig:sp_rxj}}
\end{figure}

In previous work \citep{hoetal07}, we found that our models of a
magnetic, partially ionized hydrogen
atmosphere matches well the entire spectrum, from X-rays to optical,
of \rxj\ [best-fit parameters: gravitational redshift $\zg\sim 0.22$,
magnetic field $B\approx 4\times 10^{12}$~G, and radius
$\Rinfty\approx 17$~km, where $\Rinfty=R(1+\zg)$;
see Figure~\ref{fig:sp_rxj}].
With the discovery of rotational modulation of the X-ray emission
\citep{tiengomereghetti07}, we use the light curves predicted by our
model (described in the previous section) to constrain the geometry
($\alpha,\zeta$) of \rxj\ \citep{ho07}.
We find angles of $< 6^\circ$ and $\approx 20-45^\circ$
(example fits ares shown in Figure~\ref{fig:pulse_rxj}).
This indicates that either the rotation and magnetic axes are
closely aligned or we are essentially seeing down the spin axis of the
NS (see also \citep{brajeromani02}).

%-------------------------------------------
%-------------------------------------------

\begin{figure}[htb]
\includegraphics[width=0.35\textwidth]{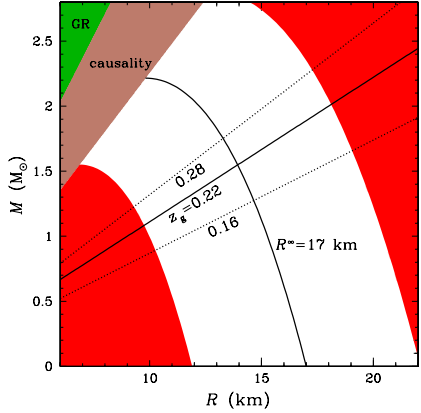}
\caption{
Constraints on NS mass $M$ and radius $R$ from fitting the
observations of \rxj.
The upper left regions are excluded by general relativity and causality.
The upper right and lower left regions exceed the $\sim 30\%$ uncertainty
in $\Rinfty$, which is dominated by the uncertainty in the distance
(see \citep{kaplanvankerkwijk07}).
The dotted lines indicate 3$\sigma$ uncertainty in $\zg$; note that this
is just the uncertainty from the fit and does not include systematic
uncertainties in the data and model.
\label{fig:eos}}
\end{figure}

\begin{figure}[htb]
\includegraphics[width=.4\textwidth]{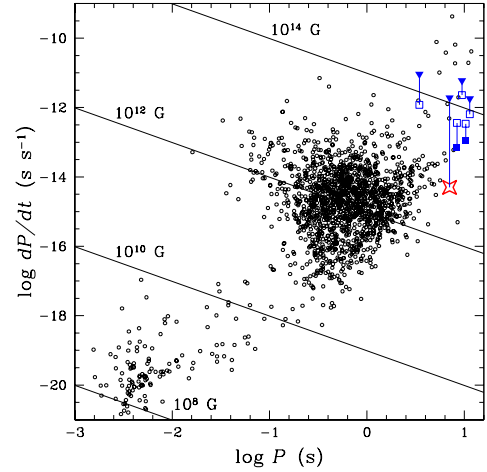}
\caption{
NS period derivative $\dot{P}$ as a function of spin period $P$.
Lines indicate constant magnetic field
[$B=3.2\times 10^{19}(P\dot{P})^{1/2}$~G].
Dots are pulsars whose $P$ and $\dot{P}$ are obtained from
{http://www.atnf.csiro.au/research/pulsar/psrcat/}
(see \citep{manchesteretal05}).
We highlight a particular class of NSs, for which \rxj\ is a member:
sources with a measured $B$ from their X-ray spectrum (open squares)
and the same sources with a measured $\dot{P}$ (solid squares) or upper
limit to $\dot{P}$ (triangles) (see \citep{haberl07}, and references therein).
The star indicates \rxj\ from our measured $B$.
\label{fig:ppdot}}
\end{figure}

From the results of our modeling of the phase-resolved observations,
we can better determine where \rxj\ belongs in the broader context
of NS studies.
For example, Figure~\ref{fig:eos} shows the constraints placed on the
mass-radius relationship for NSs.  Our results imply a relatively stiff
but standard nuclear EOS (see, e.g., \citep{lattimerprakash07}).
Also if \rxj\ is losing rotational energy by magnetic dipole
radiation, the rate of spindown is given by
$dP/dt=10^{-15}\mbox{ s s$^{-1}$}(B/\mbox{10$^{12}$ G})^2
(P/\mbox{1 s})^{-1}=5\times 10^{-15}\mbox{ s s$^{-1}$}$;
this is well below the upper limit
$dP/dt < 1.9\times  10^{-12}\mbox{ s s$^{-1}$}$ obtained by
\citep{tiengomereghetti07} and illustrated in Figure~\ref{fig:ppdot}.
Note that \rxj\ has not been detected in the radio
\citep{brazierjohnston99,burgayetal07,kondratievetal07}, though its
location in $P-\dot{P}$ space (see Figure~\ref{fig:ppdot}) is below
the theoretical death line for radio pulsars.

%-------------------------------------------
%-------------------------------------------

%-------------------------------------------
%-------------------------------------------

%%%%%%%%%%%%%%%%%%%%%%%%%%%%%%%%%%%%%%%%%%%%%%%%%%%%%%%%%
\section{\onee \label{sec:1e1207}}

\onee\ is a NS (with spin period = 0.424~s) in the center of supernova
remnant G~296.5+10.0.
Its X-ray spectrum is remarkable in that it shows two broad
absorption features at $\sim 0.7$ and $\sim 1.4$~keV
\citep{mereghettietal02,sanwaletal02,morietal05,delucaetal07,
gotthelfhalpern07b}.
Even more surprising is that these features show greater phase-variability
than the continuum spectrum \citep{bignamietal03,delucaetal04}.
Proposed models involving ion cyclotron or atomic lines from a light
element atmosphere at $B\sim 10^{14}$~G
\citep{sanwaletal02,turbinerlopez04}
seem unlikely due to weakening of line strengths
by vacuum resonance effects \citep{holai03,holai04,vanadelsberglai06}
or low abundance of the ionization states possibly responsible for the
observed lines \citep{morihailey06}.
Recent timing analysis also imply $B<3.3\times 10^{11}$~G
(by assuming vacuum dipole braking);
this suggests the spectral features are electron cyclotron lines
\citep{gotthelfhalpern07}.
Alternatively, \citep{haileymori02,morihailey06}
proposed a mid-$Z$ element atmosphere at $B\sim 10^{12}$~G.
We are studying this last case by building atmosphere models
composed of mid-$Z$ elements (see \citep{moriho07}, for details)
in an attempt to fit the phase-resolved observations.

\begin{figure}[htb]
\includegraphics[width=.4\textwidth]{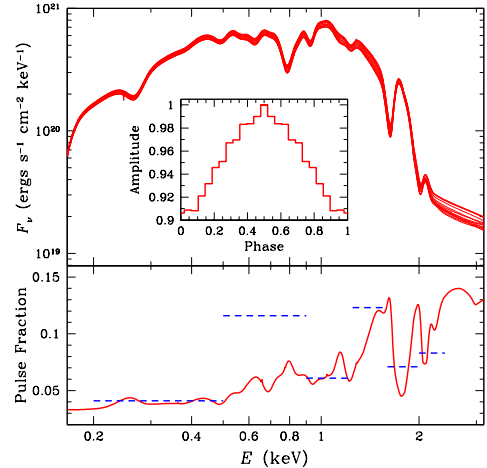}
\caption{
Top: Model atmosphere spectra (convolved with the \xmm\ EPIC-pn energy
resolution) at different rotation phases.
Inset: Energy-integrated light curve.
Bottom: Pulse fraction as a function of energy.
The dashed horizontal lines indicate the observed pulse fractions
over the given energy range for \onee\
(see \citep{delucaetal04}).
\label{fig:model1207}}
\end{figure}

As an illustration of our models, Figure~\ref{fig:model1207} shows the
phase-resolved model oxygen atmosphere spectra, light curve, and pulse
fraction [$=(C_{\rm max}-C_{\rm min})/(C_{\rm max}+C_{\rm min})$,
where $C$ is the count spectrum].
We assume here that $\zg=0.2$, $(\alpha,\zeta) = (10^\circ,20^\circ)$,
and the NS surface is parametrized by
$\Teff=4\times10^6$~K and [$B(10^{12}\mbox{G}),\ThetaB$]= ($1.2,0^\circ$)
for magnetic colatitudes 0--5$^\circ$
and ($1,30^\circ$) for colatitudes 6--20$^\circ$.
Note that our models roughly match the continuum shape and line
locations and strengths of the phase-averaged spectrum of \onee.
More detailed analysis will be reported elsewhere.

%-------------------------------------------
%-------------------------------------------

%%%%%%%%%%%%%%%%%%%%%%%%%%%%%%%%%%%%%%%%%%%%%%%%%%%%%%%%%
% \section{}

In summary, we discussed briefly our continuing work on understanding the
emission process of NS surfaces, as well as detailed comparisons
of our models to phase-resolved observations of the NSs \rxj\ and \onee.
Analyses of other sources is ongoing.  In the near future, we are
integrating our model spectra into \texttt{XSPEC} for community use
\citep{hoetal07b}.

%%%%%%%%%%%%%%%%%%%%%%%%%%%%%%%%%%%%%%%%%%%%%%%%%%%%%%%%%
\begin{theacknowledgments}
WH is extremely grateful to Gilles Chabrier, Philip Chang, David Kaplan,
Dong Lai, Alexander Potekhin, and Matthew van Adelsberg, without
whom this work would not have been possible.
WH appreciates the use of the computer facilities at the Kavli
Institute for Particle Astrophysics and Cosmology.
\end{theacknowledgments}

%%%%%%%%%%%%%%%%%%%%%%%%%%%%%%%%%%%%%%%%%%%%%%%%%%%%%%%%%
\bibliographystyle{aipproc}   % if natbib is available

\end{document}